\documentclass[twocolumn,twoside,slac_two]{revtex4}
\usepackage{graphicx}
\usepackage{fancyhdr}
\newcommand{\be}{\begin{equation}}
\newcommand{\ee}{\end{equation}}
\newcommand{\ba}{\begin{eqnarray}}
\newcommand{\ea}{\end{eqnarray}}
\begin{document}

\onecolumngrid
\begin{flushright}
    LU TP 06-37\\
    hep-lat/0610083\\
    October 2006
\end{flushright}

\title{Quark anti-quark expectation value in finite volume }

%

\author{Johan Bijnens}
\author{Karim Ghorbani\footnote{Speaker}}
\affiliation{Department of Theoretical Physics, Lund University, S\"olvegatan 14A, SE 223 62, Sweden}

\begin{abstract}
We have computed the quark anti-quark expectation value in finite volume at two loop 
in chiral perturbation theory and compare it with a formula obtained in analogy to 
the L\"uscher formula for pion mass in finite volume. We observe 
that due to the small finite size correction at two loop it is not possible to 
obtain conclusions on the accuracy of the extended L\"uscher formula.

\end{abstract}

\maketitle

\thispagestyle{fancy}


\section{Introduction}

Quantum chromo dynamics (QCD) is a strongly coupled theory at low energy.
Therefore it is difficult to study the physics below the QCD scale of about 1~GeV. 
Lattice QCD on the other hand is an attempt to tackle this problem by taking the 
Lagrangian and calculating (parts of) the generating functional numerically. 
Although these computations with fairly small quark masses are now possible, 
at the same time the finite volume effects are also becoming important. 
In this work we employ chiral perturbation theory (ChPT) in order to study these
effects analytically. Chiral perturbation theory, in its modern form,
was first proposed by 
Weinberg \cite{Weinberg} and extended in \cite{GL1,BCE}. ChPT is an effective field 
theory to describe the strong interactions at low energy. To do the 
computation in finite volume as suggested in \cite{GL2} one 
imposes a periodic boundary condition on fields which result in 
the momentum quantization and consequently modification on the quantum 
corrections. Since the applicability of ChPT is limited to small momenta, in 
finite volume this leads to 
\be
F_{\pi} L > 1 \,. 
\ee
which holds for L larger than about 2 fm.  
$F_{\pi}$ is the pion decay constant and L is the linear 
size of the boundary. In addition we stay in a limit where the 
Compton wavelength of the pion is smaller than the lattice size 
which corresponds to 
\be
m_{\pi}^2 F_{\pi}^2 V >>1 \,.
\ee 
This is the so-called p-regime.
This condition ensures that perturbative treatment will work. 
Expectation value of the quark condensate \emph{per se} will be achieved by varying 
the generating functional with respect to the scalar external field as follows 
\be 
<\overline{q}T^{a}q > =  \frac{1}{Z_{0}} (-i \frac{\delta}{\delta S_{a}} ) 
Z_{eff}[s_{a}]_{ \arrowvert s_{a}=0}.
\ee
Where $T^{a}$ stands for the Gell-Mann matrices. We then carry out 
the $\langle \overline{q}q \rangle$ at two loop in finite volume within 
the framework of chiral perturbation theory. This quantity has already been evaluated at 
one loop order in \cite{DG}. Another way for the computation of the finite volume 
effect is the L\"uscher approach. In this method one obtains the leading finite size effects 
for the pion mass to all order in perturbation theory from the scattering 
amplitude \cite{Luscher1,Luscher2}. It has later been extended to evaluate the volume dependence 
of the pion decay constant\cite{CH1}. 
As the second goal we shall then obtain a formula for the the quark condensate volume 
dependence by following the same line of reasoning as in the L\"uscher work. 
This work was published in Ref.~\cite{BGH}.

\section{A L\"uscher formula for the vacuum condensate}

We were inspired by the work of L\"uscher, where he showed how the finite size effect on  
the masses of spin-less particles are related to the forward elastic scattering amplitude 
in infinite volume. This occurs when particles are enclosed in a lattice box 
and as considered in this letter with temporal direction of the space-time extended to infinity. 
To see explicitly where these effects stem from, one looks at the two-point correlation function 
with fields subjected to a periodic boundary condition. The correlator takes on the form \\[.2cm]  
\be
<\phi(x)\phi(0>_{L}  = \int \frac{dp_{0}}{2\pi} \sum_{\vec{p}}
 L^{-d} e^{-ip.x} G_{L}^{-1}(\vec{p}^{2},m_{0}^{2})
\ee
where the inverse propagator reads  
\be
G_{L}(\vec{p}^{2},m_{0}^{2}) = p^{2}-m_{0} -\Sigma_{L}(p^2)
\ee   
and $\Sigma_{L}(p^2)$ denotes the sum of the one-particle irreducible diagrams 
to be evaluated in momentum space with discretized momenta and $\vec{p} = \frac{2\pi \vec{n}}{L}$,
with $\vec{n}$ a three dimensional vector of integers.
In practice the desired quantity to compute is the change in $\Sigma(p^2)$ due to 
the finite volume correction and the difficult task in L\"uscher's work was to show that 
this consists of all self-energy diagrams with just one propagator in each diagram allowed 
to wrap around the whole position space and therefore experiencing the boundary. 
Using the Poisson summation formula the modified propagator can be obtained in terms of the 
propagator in the infinite volume accompanied by an exponential factor as follows 
\be
G_{L}(q) = \sum_{\vec{n}} e^{-iq_{i}n_{i}L_{i}} G_{\infty} 
\ee
where in three dimensions the length squared for a given 
$\vec{n}$ reads $k= \vec{n}^{2} = n_{1}^{2}+n_{2}^{2}+n_{2}^{2} $.  
For $\vec{n} = 0 $ it gives back the propagator in infinite volume. 
The crucial point comes about in Euclidean space where $q_{0} \to -iq_{0} $ 
and, as shown below, in our result for $ \langle \bar{q} q \rangle$
the exponential factor falls off rapidly 
for large $\vec{n}^2$.  L\"uscher in his analysis kept only $\vec{n}^{2} = 1$
but summation over all values shall be carried out numerically in our formula
as suggested in \cite{CH2}. 

\begin{figure}[t]
\includegraphics[width=80mm]{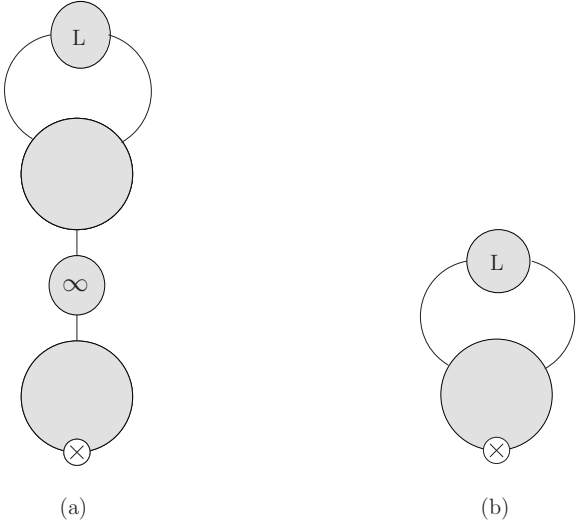}
\caption{Two potentially relevant graphs to the $\langle \bar q q \rangle$. \\ $\otimes$ indicates 
        the insertion of $\langle \bar q q \rangle$ and the solid line is the
         meson propagator \cite{BGH}.} 
         \label{JACpic2-f1}
\end{figure}

Out of three self-energy graphs shown in Ref.~\cite{Luscher1}, only two are potentially relevant 
to the $\langle \bar q q \rangle$ as depicted in Fig.~\ref{JACpic2-f1}. 
The diagram (a) has however no contribution when we apply
ChPT to the case of parity even operators. 
Extending our task to the evaluation of a general operator in the the diagram (b) 
we arrive at the following relation
\be
\langle O \rangle_{V} - \langle O \rangle_{\infty}  = \int \frac {d^{d}q} {(2\pi)^{d}}
\sum_{\vec{n} \neq 0} e^{-iL q.n} G_{\infty} \langle \phi|O|\phi \rangle \,.
\ee

Integration over momentum can be split into parallel and transverse components 
with respect to the $\vec{n}$, followed by the continuation 
$q_{\parallel} \to  q_{\parallel} -is $. 
For $s \to \infty$ the integration over this path is negligible.
Unlike the mass case it turns out that the computation of the diagram (b) 
for the  $\langle \bar q q \rangle $ is simpler mainly because the matrix 
element $\langle \phi|O|\phi \rangle $ representing the amputated 
vertex function has no external momenta in it and in fact it should be 
evaluated at zero momentum transfer.  
After isolating the poles and picking out the right one we, complete the integration
over $q_{\parallel} $ which finally leads to the following formula  
\ba
\lefteqn{\langle O_{V} \rangle - \langle O_{\infty} \rangle = - \langle \phi|O|\phi \rangle}
\nonumber\\&& 
\!\!\!\! \times\left(\sum_{\vec{n} \neq 0}  
\frac{1}{16\pi^2} \int_{0}^{\infty} \frac{q^2 dq}{\sqrt{q^2+m^2}} 
e^{-\sqrt{\vec{n}^2(m^2+q^2)L^2}} \right)
\ea             
As the last step we identify the remaining integral as the modified Bessel function 
$K_{1}$  

\be
\langle O \rangle_{V} - \langle O \rangle_{\infty} = -  \langle \phi|O|\phi \rangle 
\sum_{k=1,\infty} 
\frac{x(k)}{16\pi^2} \frac{m^2}{\sqrt{\zeta(k)}} K_{1}(\zeta(k))
\ee  
where $\zeta(k) = \sqrt(k)m L$ and $x(k)$ is the multiplicity,i.e. $x(k)$
is the number of times the value $k=\vec{n}^2$ appears in the sum overt $\vec{n}$.
Deriving this formula also has the advantage that one can estimate the size of the 
sigma term by evaluating the volume dependence of the
the quark anti-quark expectation value.
 
\section{The vacuum expectation value  at two loop  }

\begin{figure}[t]
\includegraphics[width=80mm]{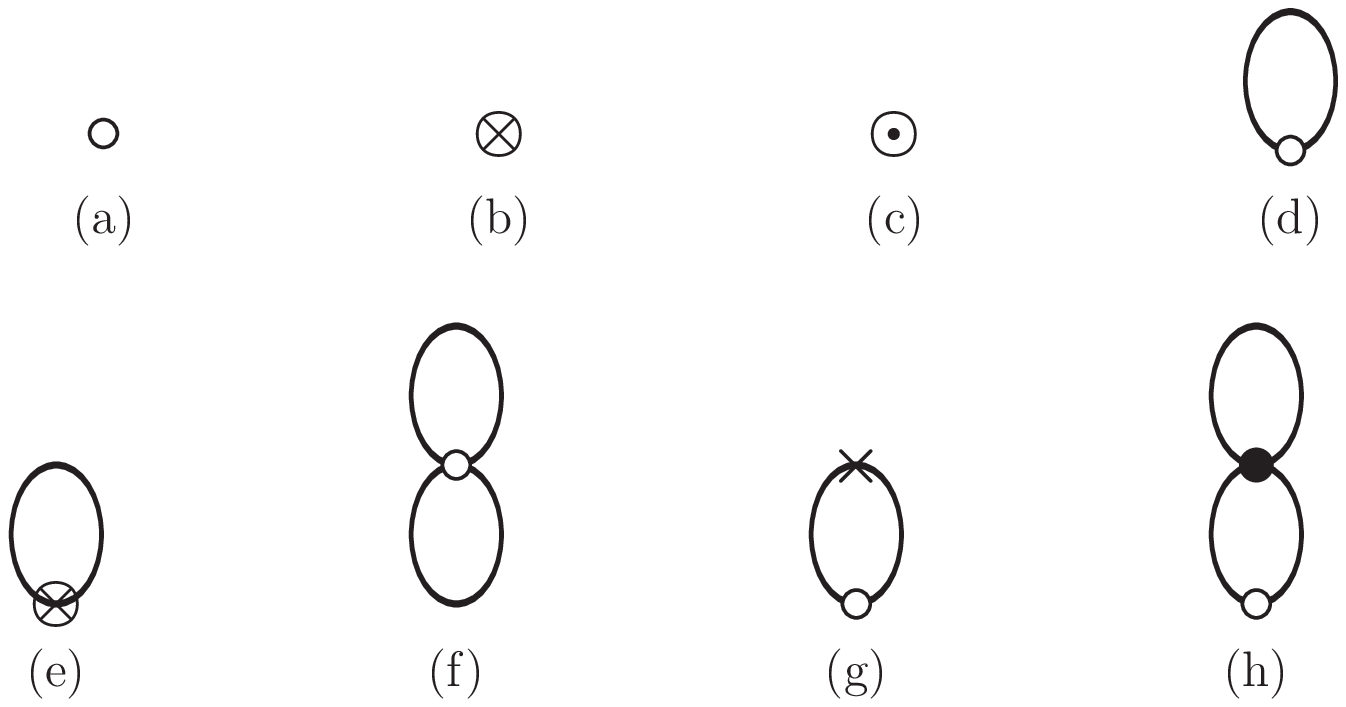}
\caption{Feynman diagrams up to $p^6$ for $\langle \bar q q \rangle$. \\ 
         vertices $\circ$, $\odot$ and $\otimes$ stand for the insertion of 
          $\langle \bar q q \rangle$
         at $p^2$, $p^2$ and $p^6$ respectively. $\bullet$ is the $p^2$ and $\times$
         the $p^4$ vertices. Solid lines indicates the meson propagator. \cite{BGH} } 
         \label{JACpic2-f2}
\end{figure}

The calculation of the quark condensate at two loop was done in \cite{ABP}. We go back to  
these computation while taking into account the finite size effects. 
As mentioned before, this effect comes about as a modification in the 
propagator. In Fig.~\ref{JACpic2-f2} we show all the Feynman diagrams that contribute up to $p^6$.
In infinite volume the loop diagrams we deal with, contain the functions

\ba
A(m^2) = \frac{i}{(2\pi)^d} \int d^dq \frac{1}{q^2-m^2} 
\nonumber \\
B(m^2) = \frac{i}{(2\pi)^d} \int d^dq \frac{1}{(q^2-m^2)^2} 
\ea

These integrals are solved in the dimensional regularization scheme 
with $\epsilon = (4-d)/2 $ .

\ba
\label{Ab}
A(m^2) &=& \frac{m^2}{16\pi^2} \left(\frac{1}{\epsilon}-\gamma_{E}+log(4\pi)+1\right)
\nonumber\\&& + \bar{A}(m^2) 
+ \epsilon A^{\epsilon}(m^2) + o(\epsilon^2 ) \,
\\
\label{Bb}
B(m^2) &=& \frac{m^2}{16\pi^2} \left(\frac{1}{\epsilon}-\gamma_{E}+log(4\pi)+1\right)
\nonumber\\&& 
+\bar{B}(m^2)+\epsilon B^{\epsilon}(m^2)+o(\epsilon^2).
\ea

At finite volume integration over momenta should be replaced by a summation
over discretized momenta. It is also important to note that the pole structure 
does not change its form, since the singularity arises when momenta in the 
loop become very large and therefore the pole remains unchanged despite the replacement 
of integration with summation. The final result for the integrals
at finite volume becomes

\ba
\label{AbV}
\bar{A}(m^2) &=& -\frac{m^2}{16\pi^2} \log\left(\frac{m^2}{\mu^2}\right)
\nonumber\\&&
 - \frac{1}{16\pi^2} \sum_{k=1,\infty}
x(k) \frac{4m^2}{\zeta(k)} K_{1}(\zeta(k))
\\
\label{BbV}
\bar{B}(m^2) &=& - \frac{1}{16\pi^2}\log\left(\frac{m^2}{\mu^2}\right)
-\frac{1}{16\pi^2} 
\nonumber\\&& 
+\frac{1}{16\pi^2} \sum_{k=1,\infty} x(k) 2K_{o}(\zeta(k))  
\ea 
with $\zeta(k) = \sqrt{k}mL$.
It is evident from equations (\ref{AbV}) and (\ref{BbV}) that the simple relation 
$\bar{A}(m^{2}) = m^2 (1/16\pi^2 +\bar{B}(m^{2}) )$ is no longer valid in 
finite volume computations. After putting all the diagrams together and 
applying the dimensional regularization scheme with renormalization 
parameter $\mu$ we ensure that the divergent parts cancel and for finite part 
we use the expressions in (\ref{AbV}) and (\ref{BbV}) instead of those in
 (\ref{Ab}) and (\ref{Bb}).  
The parts containing $A^\epsilon$ and $B^\epsilon$ cancel in the final
result for the vacuum expectation value.
We denote the lowest order masses for pions, kaons and etas as $\chi_{\pi}$, $\chi_{K}$ and
$\chi_{\eta}$ respectively. They are given in terms of the strange quark mass, $m_{s}$
and the average of up quark and down quark masses, $\hat m $ by
\ba
\chi_{\pi} = 2 B_{0} \hat m \, , \quad \chi_{K} = B_{0}(m_{s}+\hat m ) \,  
\nonumber \\
\chi_{\eta}=2B_{0}(2m_{s}+\hat m)/3
\,.
\ea
We also define the following quantities 
\ba
\langle \bar{u} u \rangle &=& -B_{0}F_{0}^2 \left(1+\frac{v_{4}^{(u)}}{F_{0}}+\frac{v_{6}^{(u)}}{F_{0}^{2}}\right) \,,
\nonumber \\
\langle \bar{s} s \rangle &=& -B_{0}F_{0}^2 \left(1+\frac{v_{4}^{(s)}}{F_{0}}+\frac{v_{6}^{(s)}}{F_{0}^{2}}  \right)\,.
\ea

\begin{figure}[t]
\includegraphics[width=80mm]{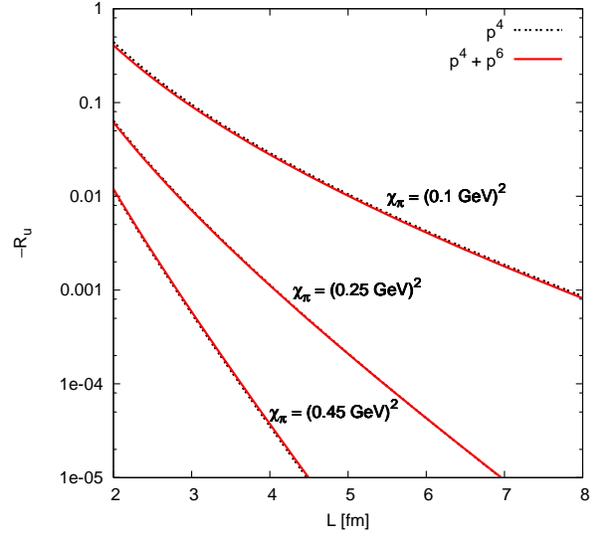}
\caption{The ratio $R_{u}$ is shown for three different pion masses\cite{BGH}. } 
         \label{JACpic2-f3}
\end{figure}

\begin{figure}[t]
\includegraphics[width=80mm]{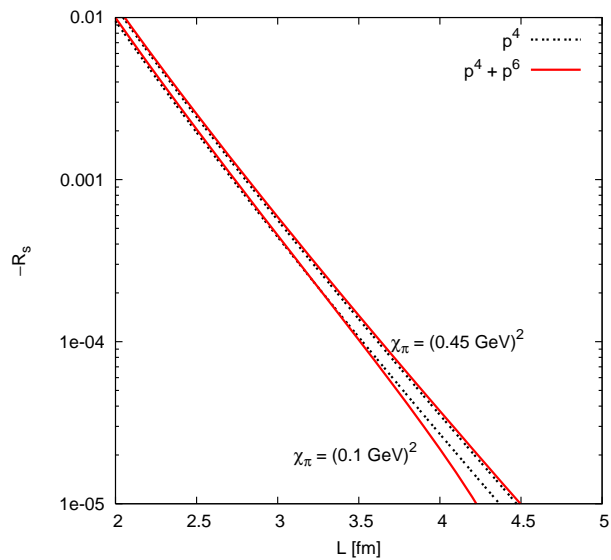}
\caption{The ration $R_{s}$ is shown for three different pion masses increasing from top 
to bottom \cite{BGH}.} 
         \label{JACpic2-f4}
\end{figure}

The analytical expressions for $ v_{4}^{(u)}$,$v_{4}^{(s)}$,
$v_{6}^{(u)}$,$v_{6}^{(s)}$ 
are evaluated in ChPT and can be found in \cite{BGH}. Numerical results are shown 
in Fig.~\ref{JACpic2-f3} and Fig.~\ref{JACpic2-f4}, where the ratio $R_{q}$ is defined as
\be
R_q \equiv\frac { \langle \bar q q \rangle_{V} - \langle \bar q q \rangle_{\infty}}
{\langle \bar q q \rangle_{\infty}} \,.
\ee
The $C^{r}_{i}$, the low energy coupling constants (LECs) from diagrams at $p^6$ \cite{BCE} 
cancel out in the numerator and we have ignored them in the denominator but the
$L^{r}_{i}$ will be considered as inputs and we use the values from fit 10 of\cite{ABT}.
As expected the volume size effects are pronounced for small L and small pion mass.
But the full NNLO contribution to the finite volume effect
is much smaller than the total NNLO numerical correction.
In turn this total NNLO correction is smaller than the one
presented in \cite{ABP}. This is due to the fact that in Ref.~\cite{ABP}
the physical value of pion decay constant $F_{\pi}$ and also physical masses for
mesons are put in the formula.

\section{Conclusions}

We compared the quark condensate in finite volume obtained in two different ways. 
The method referred to as L\"uscher's approach yields the leading term of this quantity 
to all orders in perturbation theory while only one propagator experiences the finite 
volume. One the other hand, we try to do this computation up to $p^6$ rather directly 
by letting the  modification of all propagators be present at the same time.
The comparison of the direct calculation at one-loop with the extended formula is 
trivial but at two-loop level, however, this is not the case.
This allows in principle a check of the accuracy of the extended L\"uscher approach.  
Unfortunately, because of the very small size of the finite volume effect at $p^6$,
no conclusion can be drawn concerning the accuracy level of the extended L\"uscher 
formula from our work.

\begin{acknowledgments}
One of us, KG, wishes to thank the organizers of the IPM-LHP06 conference for the travel grant and 
great hospitality.
This work is supported by the European Union RTN network,
Contract No. 
MRTN-CT-2006-035482  (FLAVIAnet) and by 
the European Community-Research Infrastructure
Activity Contract No. RII3-CT-2004-506078 (HadronPhysics).
KG acknowledges a fellowship from the Iranian Ministry of Science.

\end{acknowledgments}

\bigskip 

\end{document}